# Photonic quantum-corral ring laser: A fermionic phase transition

O'Dae Kwon[*], Byeonghoon Park[*], Junyeon Kim[*], Joongwoo Bae[*] & Moojin Kim[*]

Jungchak Ahn [†] & Oh-Hyun Kwon [†]

[*]*Elec.Eng.Dept., Pohang University of Science & Technology, San 31 Hyojadong, Pohang 790-784, Korea.*

[†] *Samsung Electronics Co., Nongseo-Ri, Ki-Heung, Yong-In 440-600, Korea.*

**Extensive Bose-Einstein condensation research activities[1,2] have recently led to studies of fermionic atoms and optical confinements[3-6]. Here we present a case of micro-optical fermionic electron phase transition. Optically confined ordering and phase transitions of a fermionic cloud in dynamic steady state are associated with Rayleigh emissions from photonic quantum ring manifold which are generated by nature without any ring lithography[7]. The whispering gallery modes, produced in a semiconductor Rayleigh-Fabry-Perot toroidal cavity at room temperature, exhibit novel properties of ultralow thresholds open to nano-ampere regime, thermal stabilities from $\sqrt{T}$ -dependent spectral shift, and angularly varying intermode spacings. The photonic quantum ring phenomena are associated with a photonic field-driven phase transition of quantum-well-to-quantum-wire and hence the photonic (non-de Broglie) quantum corral effect[8] on the Rayleigh cavity-confined carriers in dynamic steady state. Based upon the intra-cavity fermionic condensation we also offer a prospect for an electrically driven few-quantum dot**

single photon source from the photonic quantum ring laser for quantum information processors[9,10].

Current fermionic cloud shown in Fig. 1a involves a site ordering from a random distribution of two-dimensional (2D) quantum well (QW) electrons to one-dimensional (1D) quantum wire electron arrays, or as explained later to a 1D chain of quasi quantum dot (QD) electrons. To address the fermionic condensation problem we briefly explain the photonic quantum ring (PQR) phenomena here. First of all, the ultralow threshold current data at room temperature of the three-dimensional (3D) PQR laser (Fig. 1b) follow a simplified formula derived for 2D concentric quantum ring arrays[7]:

$$I_{th}^{PQR} = I_{tr}^{PQR} + I_{i}^{PQR} + I_{mirror}^{PQR} = \left(\frac{\pi n_{eff} e N_{tr}^{1D} w}{\lambda_{PQR} \eta \tau}\right)\phi^2 + \left(\frac{\pi d e \alpha_i w}{2g'_{1d}\tau}\right)\phi^2 + \left(\frac{dew \ln R^{-1}}{2g'_{1d}\tau}\right)\phi,$$

where $n_{eff}$ is the effective azimuthal refractive index, $N^{1D}_{tr}$ the transparency carrier density, $\lambda_{PQR}$ the emission wavelength, $\tau$ the carrier lifetime, $\eta$ the quantum efficiency, $w \equiv 1 - n_{eff}/n = W_{Rayleigh}/(\phi/2)$ the normalized Rayleigh bandwidth, $\phi$ the diameter, $d = \lambda_0/n$ the active region thickness, $\alpha_i$ the intrinsic loss, $g'_{1d}$ the differential gain, and $R \approx 1$.

The reason why the simplified 2D formula works well is that the actual gain comes from the 2D QW plane. Data in Fig. 1c, curve A, for either full or hollow mesa PQRs, are consistent with the theoretical prediction. It also indicates that the PQRs of 5 μm or less in diameter would begin to enter the nano-ampere threshold regime easily. Details of the device and measurements were reported earlier[7].

If the physical bandwidth of the annular PQR region occupying the Rayleigh-Fabry-Perot cavity, given by $W_{Rayleigh} = (\phi/2)(1 - n_{eff}/n)$, is regarded as the regular QW plane, the QW threshold formula is given as[11]:

$$I_{th}^{QW} = \left(\frac{\pi e N_{tr}^{2D} w}{2\eta\tau}\right)\phi^2 + \left(\frac{\pi d e \alpha_i w}{2g'_{2d}\tau}\right)\phi^2 + \left(\frac{dew\ln R^{-1}}{2g'_{2d}\tau}\right)\phi$$

One then gets curve B in Fig. 1c, being far off from the actual threshold data, which implies that the emission of the PQR is associated not with the QWs but with the quantum wires or rings in character. This difference is the reason why the threshold of the conventional vertical cavity surface emitting laser (VCSEL) is roughly up to 1,000 times that of the PQR in similar shape, as shown previously[7].

The PQR production described above implies a sudden dynamic phase transition that is induced by the PQR emission in the cavity since the Rayleigh region was simply a peripheral region of the normal QW plane before the emission. The random distribution of electrons in the QW plane must have experienced an abrupt ordering of quantum wire type for the PQR. It is abrupt since the PQR modulation behaviors at reasonably high frequencies are not so different from the VCSEL dynamics in general. The PQR threshold analysis further demands the PQR spacing to be λ/2.

A possible explanation for the fermionic condensation comes as follows: Electronic quantum corral work[8] suggests a local density of state (LDOS) along the peripheral Rayleigh toroid in the first place. Onset of the emission, even with a small injection, then tends to laser-trap and order the carriers in the cavity. Of course the fermionic condensation is for the average carriers in dynamic steady state since the e-h recombination process has already begun. The laser trapping taking place in the Rayleigh toroid will be similar to the gradient-field-induced trapping form calculated by Letokhov and Minogin[12] in a cavity of standing modes which impose a photonic (non-de Broglie) spacing of λ/2 on the neutral particle's transverse distribution. Exact calculations for such fermionic phase transitions, if available, will tell how this kind of photonic quantum corral effect (PQCE) governs the dynamic redistribution phenomena.

This PQCE seems to be the only feasible interpretation of the unusual dynamic transition of QW-to-PQR phase. We further note that the photonic spacing of λ/2 should prevail in transverse as well as longitudinal directions obviously. In fact the chains of quasi-QD will be distributed along the longitudinal standing modes trapping the dynamic carrier cloud in the Rayleigh region of the active QW plane as shown in Fig. 1a.

Recently λ/2-spaced optical corral calculations for dielectric cavities have appeared[13] and in fact such photonic spacings have been observed through near-field optical scanning tunneling experiments on the Rayleigh whispering gallery modes from an optically pumped dielectric microcavity[14]. On the other hand, low-temperature scanning tunneling microscopy studies have revealed de Broglie structures of the LDOS in electronic Fabry-Perot resonators. This LDOS study however is limited to the surface state electrons, and presently a better probing is needed in order to observe the non-de Broglie electronic structures undergoing the fermionic condensation in the PQR cavity. The PQR structure should in fact be residing in the Rayleigh torus, which is 3-5 μm below the device surface because of the top distributed-Bragg-reflector (DBR) mirror stack sitting above the active region. Nevertheless the Kapitza-Dirac experimental results are encouraging evidence associated with the PQCE in that free electrons are diffracted due to a standing light wave[15].

Spectral narrowings with increased injection currents were observed, and as shown in Fig. 1d, the narrowest linewidth observed, for example from a 10 μm device with an optical spectrum analyzer (OSA) (HP 70951A: The resolution band width limit of the OSA = 0.8 Å), is as narrow as $\Delta\lambda_{1/2}$=0.55 Å at an injection current of 800 μA, which suggests a Q value of 15,000 or better[16]. Perfectly smooth cylindrical PQRs with the chemically assisted ion beam-etched sidewall roughness << 0.1um, being far less than emission wavelengths, will further promise even narrower and scattering-free linewidths. We note that the present result is already near the linewidth of a fine single QD, observed only through optical pumping[17]. Recently fabricated 1K PQR array shows very uniform continuous-wave emissions (Fig. 1e) at room temperature, which is expected from peculiar thermal stabilities due to $\sqrt{T}$-dependent peak shifts as discussed previously[7].

The photons helically propagate clockwise and counterclockwise in the 3D Rayleigh toroid of the PQR lasers. These clockwise and counterclockwise chiral waves always share the same helical path back and forth, and their vectorial sum gives rise to the external emissions in radial direction. We have earlier reported that this intra-cavity chiral propagation leads to a breaking of vertical-axis rotational symmetry, leading to the evanescent tunneling and propagation crossover unlike any other microcavity lasers[18].

Although the wavelengths of individual modes are predetermined for a given device, typical Rayleigh resonances also exhibit angularly varying peak envelopes[19] of the off-normal resonance as shown in Fig. 2. Another angular variation of the intermode spacings only observed in the PQR laser is also peculiar, and may become a new source for the wavelength-division-multiplexing (WDM) scheme utilizing non-equal interval spectra in the future. A family of angle vs. mode spacing curves is now compared with a knot model[20] calculation, which gives an excellent account of 3D WDM ray propagation characteristics of the PQR manifold such as the non-equal mode spacing. Details of the knot model analysis will be reported elsewhere. We notice a growing trend of the mode spacings and widening modal decompositions as the device size is reduced. Actually the number of rings given by $\chi = 2n_{eff}W_{Rayleigh}/\lambda_{PQR}$ is predicting a single PQR regime and emergence of single frequency lasing as well for $\phi < 3$ μm for current GaAs/AlGaAs PQRs.

Self-organized multiple QD lasers[21,22] have appeared in mid-90s and the QD laser work applied to the realization of few-QD single photon sources (SPS)[23,24] has only recently made substantial progress for quantum computing applications[25-27]. The current QD SPS studies however rely mostly upon multi QDs grown on the active VCSEL layer[24], the surface of 2D micro-disks used for thumbtack-type whispering gallery lasers[25], or the active layer of light-emitting diodes[26]. These methods will typically involve 'by chance' fabrication steps like QD growth, mesa etching, confirmation of the surfactant QDs, and selection of appropriate single or few QDs before identifying the right SPS dot, which will be rather tedious and highly unreliable for any effective device fabrications.

Considering a knot model illustration like Fig. 3a for the PQR's chiral manifold, what will happen when we reduce the PQR diameter size toward and below the emission wavelength? Figure 3b for instance corresponds to a wavelength-size PQR which carries only a couple of standing modes let alone the single frequency behavior, so that a novel PQR SPS laser can be realized 'at will' not 'by chance' as follows: Figure 3c shows a hyperboloid structure being made for 0.1-0.3 μm PQRs. The top surface is large enough for easy electrode fabrication while the central minimum region of the active QW plane serves as the few-QD Rayleigh cavity. Properly pinned PQR modes from a 2 QD states like Fig. 3b will give a dipole radiation, useful for the SPS operation. As the device size shrinks the surface nonradiative recombination loss becomes significant. Nevertheless it is worthwhile to test micro-to-nano-size devices in a near future since such a nano-PQR is for ultrasmall injection purpose, and the surface loss, being proportional to the carrier-density, may not harm the device for quantum dot applications.

Acknowledgements
This work was partially supported by NRL, KOSEF, POSCO, and Samsung. O'Dae Kwon thanks Profs. I. Prigogine and B. Widom for their interest.

**Correspondence and requests for materials should be addressed to O'Dae Kwon.**

**(e-mail : odkwon@postech.ac.kr).**

**Figure legends**

Figure 1. Characteristics of the photonic quantum ring laser similar to a VCSEL pillar in structure. **a**, Schematics of phase transitions **b**, Near field top view of the PQR laser (Diameter $\phi$=30 $\mu$m, I=15 $\mu$A) whose evanescent-tunneled propagation appears from the active waist region. A peripheral PQR ring is brighter than the central emission. The inset shows the appearance of central VCSEL mode at I $\geq$ 15 mA. A 28.6 dBm neutral density filter was added for the emission capture. **c**, Measured thresholds vs. the ideal threshold curves (A for PQR; B for QW) predicted from the 2D concentric quantum ring calculation. **d**, Linewidth of the fundamental mode is as narrow as 0.55 Å measured from a 10 $\mu$m device with 0.8 mA injection current. **e**, Near field photograph of a 1.6 × 1.6 $mm^2$ 1K PQR laser array taken with a total injection current of 1 mA.

Figure 2. Multi-chromatic spectra and mode spacings of the PQR lasers in different diameters. **a**, Non-equal mode spacings which get larger as $\phi$ decreases, where $\phi$ = 36 $\mu$m, 24 $\mu$m, and 10 $\mu$m with the injection currents of 6 mA, 2.5 mA, and 0.7 mA, respectively. **b**, Various symbols (solid) are for the measured data for each device. Dotted lines of empty marks represent the torus knot model calculations.

Figure 3. Single photon source implementation scheme with the PQR laser. **a**, The curved chiral wave propagation in 3D Rayleigh toroidal cavity is illustrated for the case of slant angle = 46.1° by using the knot model K(3,66). **b**, A schematic of 2 QD states using the torus knot K(1,2). **c**, A SEM micrograph of the hyperboloid shape mesa.

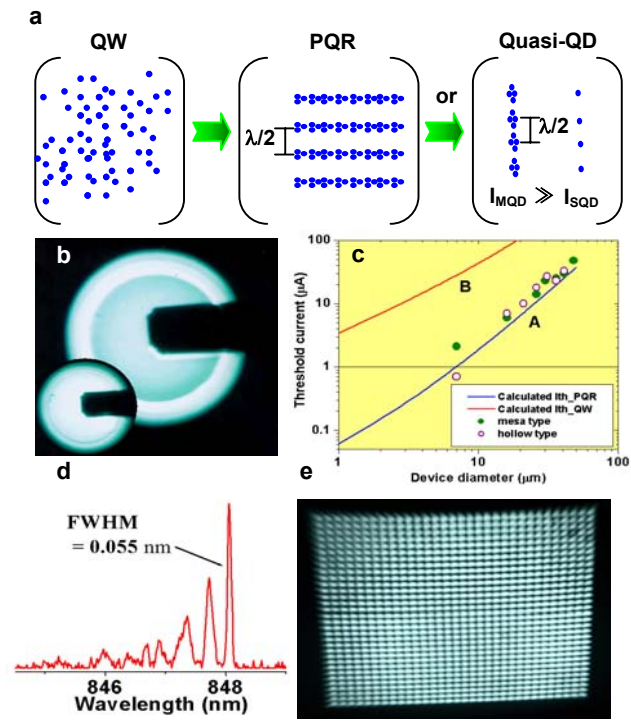

Figure 1. O'Dae Kwon

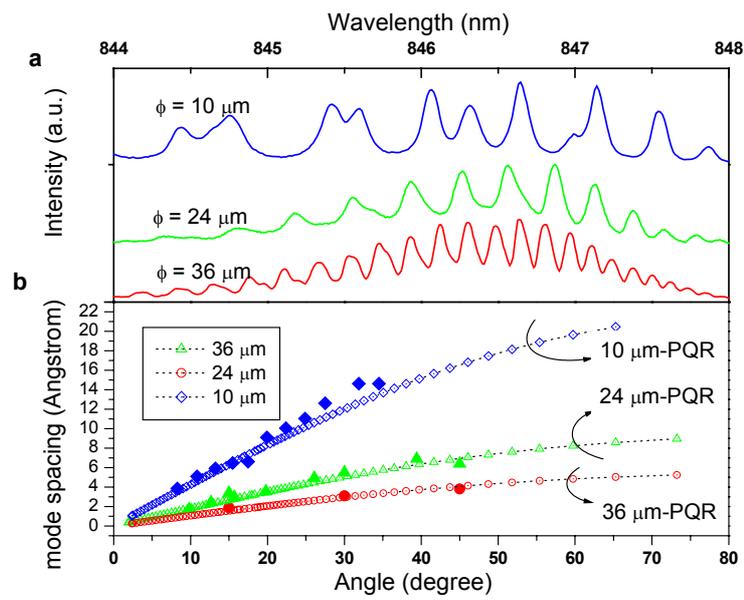

Figure 2. O'Dae Kwon

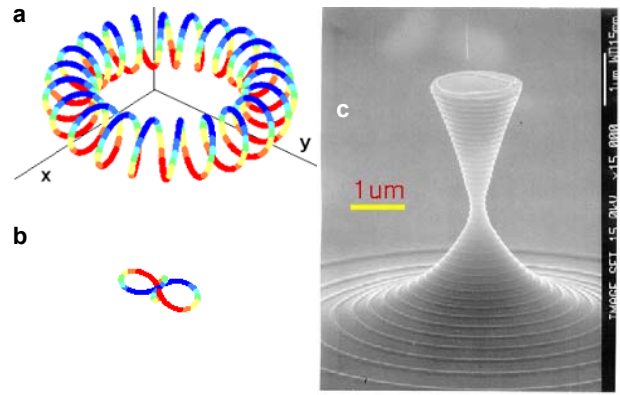

Figure 3. O'Dae Kwon